\documentclass[preprint]{aastex}
\usepackage{graphicx}

\newcommand{\teff}{\mbox{$T_{\rm eff}$}}
\newcommand{\logg}{\mbox{$\log g$}}
\newcommand{\vsini}{\mbox{$v \sin i$}}
\newcommand{\mictrb}{\mbox{$\xi_{\rm t}$}}
\newcommand{\mactrb}{\mbox{$v_{\rm mac}$}}

\newcommand{\kms}{\mbox{km\,s$^{-1}$}}
\newcommand{\halpha}{\mbox{$H_\alpha$}}

\shorttitle{WASP-32b}
\shortauthors{Maxted et~al.}

\begin{document} 
\title {WASP-32b: A  transiting ``hot Jupiter'' planet orbiting a
lithium-poor, solar-type star.}
\author{
P.F.L.~Maxted\altaffilmark{1}, 
D.R.~Anderson\altaffilmark{1}, 
A.~Collier Cameron\altaffilmark{2}, 
M.~Gillon\altaffilmark{3}, 
C.~Hellier\altaffilmark{1}, 
D.~Queloz\altaffilmark{4}, 
B.~Smalley\altaffilmark{1}, 
A.H.M.J.~Triaud\altaffilmark{4}, 
R.G.~West\altaffilmark{5},
R.~Enoch\altaffilmark{2}, 
T.A.~Lister\altaffilmark{6}, 
F.~Pepe\altaffilmark{4}, 
D.L.~Pollacco\altaffilmark{7}, 
D.~S\'egransan\altaffilmark{4}, 
I.~Skillen\altaffilmark{8}, 
S.~Udry\altaffilmark{4}}

\altaffiltext{1}{Astrophysics Group, Keele University, Staffordshire, ST5 5BG, UK}
\altaffiltext{2}{School of Physics and Astronomy, University of St. Andrews, North Haugh, Fife, KY16 9SS, UK}
\altaffiltext{3}{Institut d'Astrophysique et de G\'eophysique,  Universit\'e de Li\`ege,  All\'ee du 6 Ao\^ut, 17,  Bat.  B5C, Li\`ege 1, Belgium}
\altaffiltext{4}{Observatoire de Gen\`eve, Universit\'e de Gen\`eve, 51 Chemin des Maillettes, 1290 Sauverny, Switzerland}
\altaffiltext{5}{Department of Physics and Astronomy, University of Leicester, Leicester, LE1 7RH, UK}
\altaffiltext{6}{Las Cumbres Observatory, 6740 Cortona Dr. Suite 102, Santa Barbara, CA 93117, USA}
\altaffiltext{7}{Astrophysics Research Centre, School of Mathematics \& Physics, Queen's University, University Road, Belfast, BT7 1NN, UK}
\altaffiltext{8}{Isaac Newton Group of Telescopes, Apartado de Correos 321,
E-38700 Santa Cruz de la Palma, Tenerife, Spain}

\begin{abstract}
 We report the discovery of a transiting planet orbiting the star
TYC~2-1155-1. The star, WASP-32, is a moderately bright (V=11.3) solar-type
star ($\teff=6100\pm 100$\,K, [Fe/H]$ = -0.13\pm 0.10$). The lightcurve of the
star obtained with the WASP-South and WASP-North instruments shows periodic
transit-like features with a depth of about 1\% and a duration of 0.10\,d
every 2.72\,days. The presence of a transit-like feature in the lightcurve is
confirmed using z-band photometry obtained with  Faulkes Telescope North. High
resolution spectroscopy obtained with the CORALIE  spectrograph
confirms the presence of a planetary mass companion. From a combined analysis
of the spectroscopic and photometric data, assuming that the star is a typical
main-sequence star, we estimate that the planet has a mass $M_{\rm p}$ of
$3.60\pm 0.07M_{\rm Jup}$ and a radius $R_{\rm p} = 1.19 \pm 0.06R_{\rm Jup}$.
WASP-32 is one of a small group of ``hot Jupiters'' with masses $>  3M_{\rm
Jup}$. We find that some stars with hot Jupiter companions and with masses
$M_{\star}\approx 1.2\,M_{\sun}$, including WASP-32, are depleted in lithium,
but that the majority of these stars have similar lithium abundances to field
stars. 

\end{abstract}


\section{Introduction}
 
  The WASP project \citep{2006PASP..118.1407P} is currently one of the most
successful wide-area surveys designed to find exoplanets transiting relatively
bright stars.  Other successful surveys include HATnet
\citep{2004PASP..116..266B}, XO \citep{2005PASP..117..783M} and TrES
\citep{2006AAS...20922602O}. The Kepler satellite is now also starting to find
many transiting exoplanets \citep{2010Sci...327..977B}. There is continued
interest in finding transiting exoplanets because they can be accurately
characterized and studied in some detail, e.g., the mass and radius of the
planet can be accurately measured. This gives us the opportunity to explore
the relationships between the density of the planet and other properties of
the planetary system, e.g., the semi-major axis, the composition and spectral
type of the star, etc. Given the wide variety of transiting planets being
discovered and the large number of parameters that characterize them,
statistical studies will require a large sample of systems to identify and
quantify the relationships between these parameters. These relationships can
be used to test models of the formation, structure and evolution of short
period exoplanets. Here we report the discovery of a ``hot Jupiter'' companion
to the star WASP-32 and show that this star is lithium poor compared to other
stars of similar mass.

\section{Observations}

 The two WASP instruments each consist of an array of 8 cameras with Canon
200-mm f/1.8 lenses and  2k$\times$2k $e$2$V$ CCD detectors providing image
with a field-of-view  of $7.8^{\circ}\times 7.8^{\circ}$ at an image
scale of 13.7 arcsec/pixel \citep{2006PASP..118.1407P, 2008ApJ...675L.113W}.
The star TYC~2-1155-1 (= 1SWASP J001550.81$+$011201.5) was observed 5906 times
in one camera of the WASP-South instrument in Sutherland, South Africa during
the interval 2008 June 30 to 2008 Nov 17. Our transit detection algorithm
\citep{2007MNRAS.380.1230C} identified a periodic feature with a depth of
approximately 0.01 magnitudes recurring with a 2.72-d period in these data.
The width and depth of the  transit are consistent with the hypothesis that it
is due to a planet with a radius of approximately 1 Jupiter radius orbiting a
solar-type star. The proper motion and catalogue photometry available for
TYC~2-1155-1 suggest that it is a mid-F type, main-sequence star. We therefore
added this star to our programme of follow-up observations for candidate
planet host stars.

\begin{figure} 
\plotone{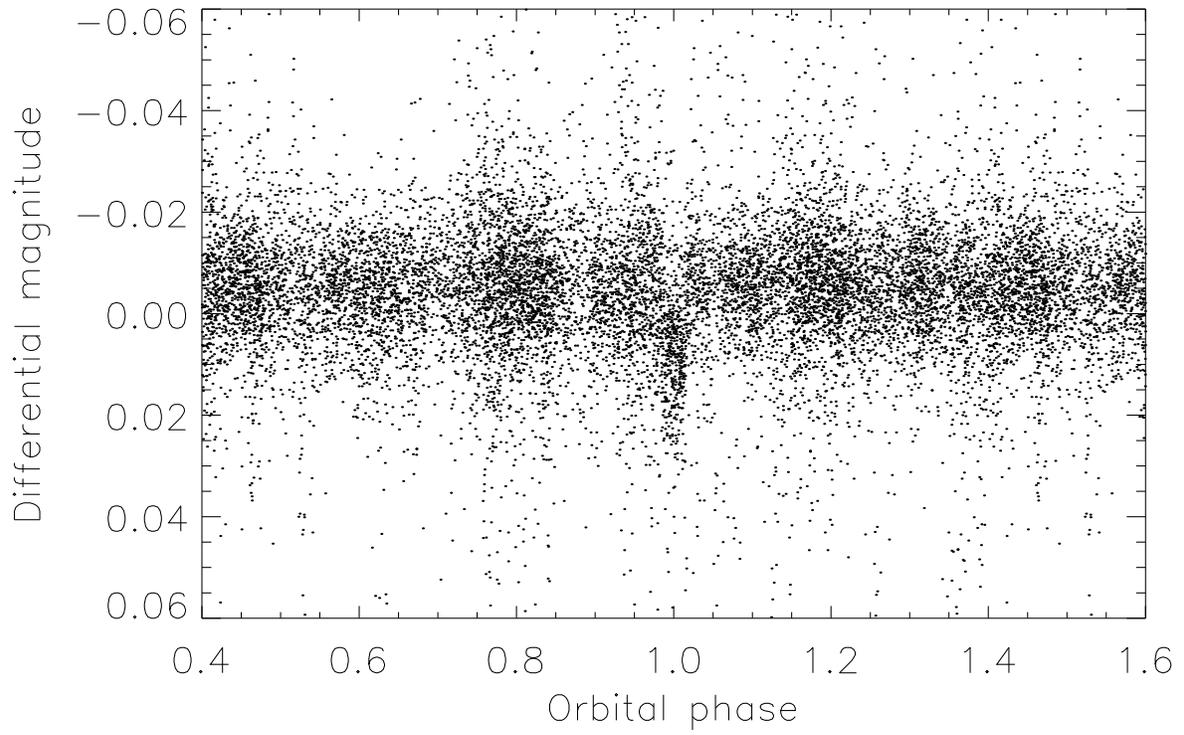} 
\caption{WASP photometry of WASP-32 folded on the orbital period
$P$~=~2.71866\,d. 
\label{wasplc} } 
\end{figure}

 A further 4156 observations of TYC~2-1155-1 were secured with WASP-South in
the interval 2009 June 28 to 2009 Nov 18. TYC~2-1155-1 was also observed by
the WASP-North instrument  2687 times during the interval 2008 August 4 to
2008 November 30 and 4308 times during the interval  2009 August 5 to 2009 Oct
20. All these data are shown as a function of phase in
Fig.~\ref{wasplc}.
\footnote{All photometric data presented are this paper are available from
the NStED database\footnote{\url{http://nsted.ipac.caltech.edu}}.}

\begin{figure} 
\plotone{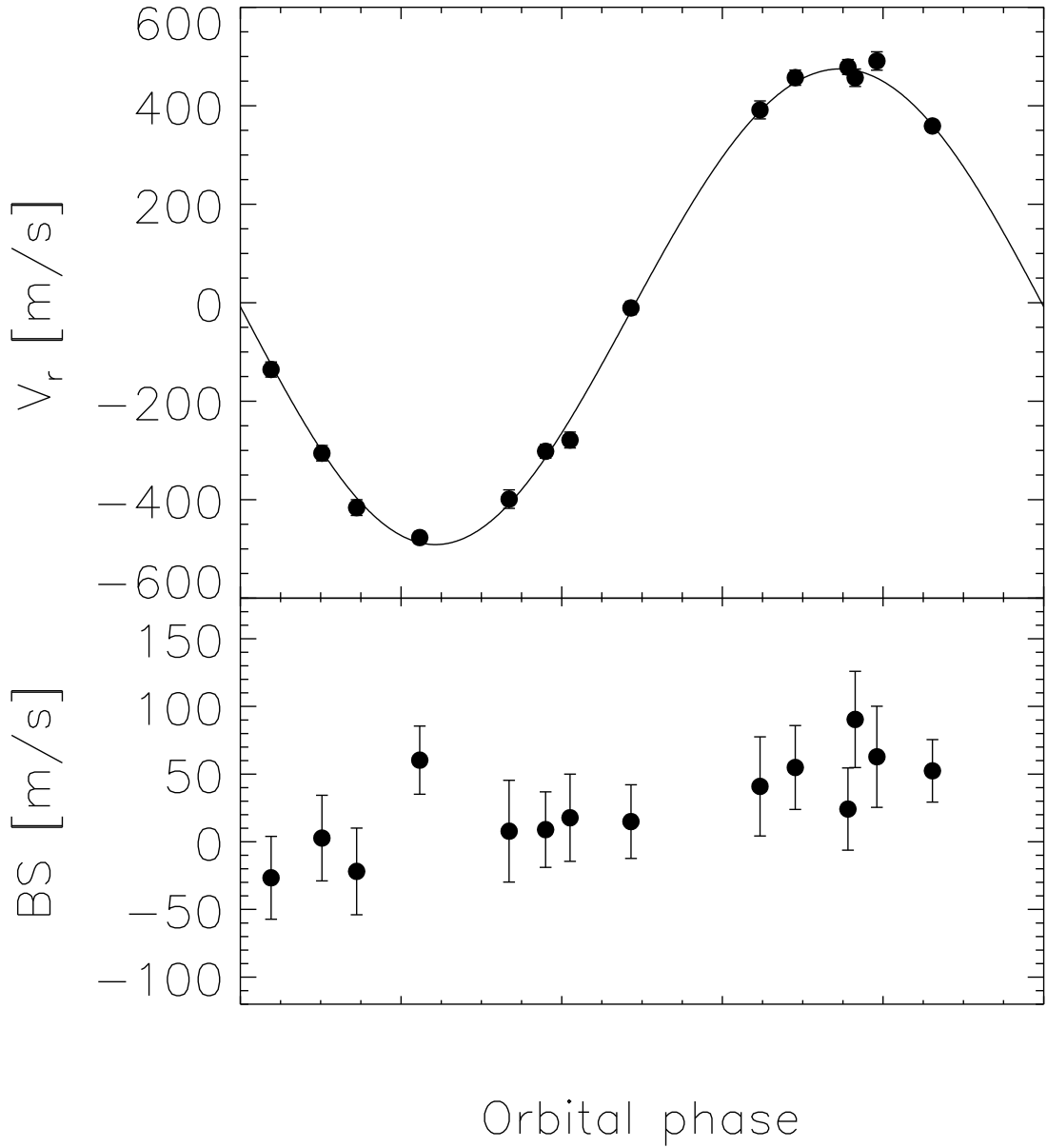} 
\caption{Radial velocity and bisector span measurements for WASP-32. Upper
panel: Radial velocity data (points with error bars) with our model for the
spectroscopic orbit (solid line). Lower panel: bisector span measurements.
(One point with large error bars is not shown here).
\label{Bis-RV} }
\end{figure}

 We obtained 15 radial velocity measurements of WASP-32 during the interval
2009 September 1 to 2009 December 22 with the CORALIE spectrograph on the
Euler 1.2-m telescope located at La Silla, Chile (Table~\ref{rv-data}). The
amplitude of the radial velocity variation with the same period as the transit
lightcurve (Fig.~\ref{Bis-RV}) and the lack of any correlation between this
variation and the bisector span establish the presence of a planetary mass
companion to this star \citep{2001A&A...379..279Q}.

\begin{figure} 
\plotone{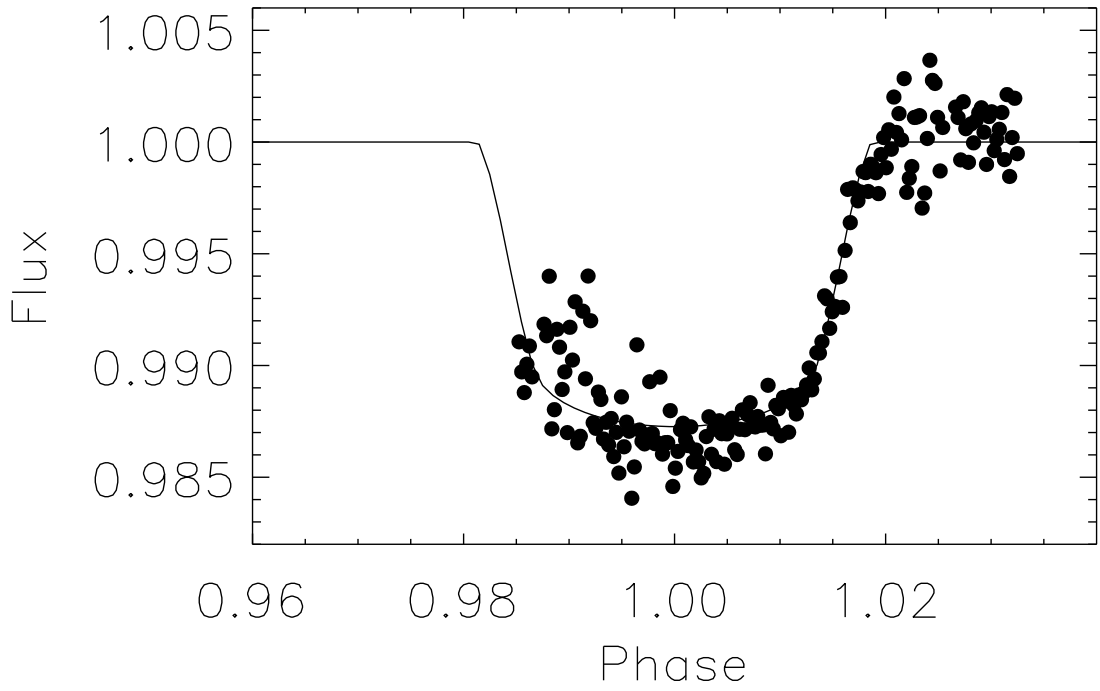} 
\caption{FTN z-band photometry of the transit of WASP-32 (points) together
with a model lightcurve for our best-fitting model parameters (solid line). 
\label{fup-phot} }
\end{figure}

 We  obtained photometry further photometry of TYC~2-1155-1 and other nearby
stars on 2009 December 7 using the fs03 Spectral camera on the LCOGT 2.0-m
Faulkes Telescope North (FTN) at Haleakala, Maui in order to better define the
depth and width of the transit signal.  The Spectral camera used a
4096$\times$4096 pixel Fairchild CCD with 15$\micron$ pixels which were binned
$2\times2$ giving an image scale of 0.303 arcsec/pixel and a field-of-view of
10'\,$\times$\,10'. We used a Pan-STARRS z filter to obtain 187 images
covering one egress of the transit. These images were pre-processed using the
WASP Pipeline \citep{2006PASP..118.1407P} using a combined bias and flat frame
and the DAOPHOT photometry package \citep{1987PASP...99..191S} was used within
the IRAF\footnote{IRAF is distributed by the National Optical Astronomy
Observatory, which is operated by the Association of Universities for Research
in Astronomy (AURA) under cooperative agreement with the National Science
Foundation.} enviroment to perform object detection and aperture photometry
with an aperture size of 9 binned pixels in radius. Differential magnitudes
were derived by combining the flux of 11 stable comparison stars that were
within the instrumental field-of-view. The resulting lightcurve is shown in
Fig.~\ref{fup-phot}. The coverage of the out-of-transit phases is quite
limited, but the data are sufficient to confirm that transit-like features
seen in the WASP data are due to the star TYC~2-1155-1 and to provide precise
measurements of the depth of the transit and the duration of egress.

\section{WASP-32 Stellar Parameters}

The 15 individual CORALIE spectra of WASP-32 were co-added to produce a
single spectrum with an approximate signal-to-noise ratio of around 80:1. The
standard pipeline reduction products were used in the analysis.

The analysis was performed using the methods given in
\cite{2009A&A...496..259G}. The \halpha\ line was used to determine the
effective temperature (\teff), while the Na {\sc i} D and Mg {\sc i} b lines
were used as surface gravity (\logg) diagnostics.  The elemental abundances
were determined from equivalent width measurements of several clean and
unblended lines.  Atomic line data was mainly taken from the
\cite{1995all..book.....K} compilation, but with updated van der Waals
broadening coefficients for lines in \cite{2000A&AS..142..467B} and $\log gf$
values from \cite{2000AJ....119..390G}, \cite{2001AJ....121..432G} or
\cite{2004A&A...415.1153S}. Individual lines abundances were determined from
the measured equivalent widths. The mean values relative to solar are given in
Table~\ref{wasp32-params}.

 A value for microturbulence (\mictrb) was determined from the Fe~{\sc i}
lines using Magain's (1984) method. The quoted error estimates include that
given by the uncertainties in \teff, \logg\ and \mictrb, as well as the
scatter due to measurement and atomic data uncertainties.

 The projected stellar rotation velocity (\vsini) was determined by fitting
the profiles of several unblended Fe~{\sc i} lines. A value for
macroturbulence (\mactrb) of 4.7 $\pm$ 0.3 \kms\ was assumed, based on the
tabulation by Gray (2008), and an instrumental FWHM of 0.11 $\pm$ 0.01\,\AA,
determined from the telluric lines around 6300\,\AA. A best fitting value of
\vsini\ = 4.8 $\pm$ 0.8~\kms\ was obtained.

\subsection{Planetary parameters}
 The CORALIE radial velocity measurements were combined with the WASP-South,
WASP-North and FTN z-band photometry in a simultaneous Markov-chain
Monte-Carlo (MCMC) analysis to find the parameters of the WASP-32 system. The
shape of the transit is not well defined by the available photometry, so we
have imposed an assumed main-sequence mass\,--\,radius relation as an additional
constraint in our analysis of the data.  The  stellar mass is determined from
the parameters \teff, \logg\ and [Fe/H] using the procedure described by
\cite{2010A&A...516A..33E}. The code uses \teff\ and [Fe/H] as MCMC jump
variables, constrained by Bayesian priors based on the
spectroscopically-determined values given in Table~\ref{wasp32-params}. The
parameters derived from our MCMC analysis are listed in
Table~\ref{sys-params}.

  Note that in constrast to previous analyses of this type, we now use the
parameters $\sqrt{e\cos{\omega}}$ and $\sqrt{e\sin{\omega}}$ in the MCMC
analysis. This removes a bias in the analysis towards larger values of $e$,
i.e., using $e\cos{\omega}$ and $e\sin{\omega}$ effectively imposes a prior
probability distribution for $e\propto e^2$, whereas using
$\sqrt{e\cos{\omega}}$ and $\sqrt{e\sin{\omega}}$ is equivalent to a uniform
prior for $e$. Nevertheless, the MCMC solution suggests a marginal detection
of a non-zero orbital eccentricity (2.8$\sigma$). This result can be confirmed
by measuring the phase of the secondary eclipse, which is expected to be
displaced from phase 0.5 by 40 minutes given our best estimate of
$e\cos{\omega}$.

 The chi-squared value of the fit to our 15 radial velocity measurements is
$\chi^2_{\rm rv}=16.7$. There are 6 free parameters; $P$ and $T_0$ are
determined almost entirely by the lightcurves, $K$ and $\gamma$ are determined
entirely by the radial velocity data but $e$ and $\omega$ are constrained by
both data sets. The number of degrees of freedom is $N_{\rm df} \approx
15-3=12$.
We did not consider it necessary to increase the standard errors of the radial
velocity data to account for  external noise due to stellar activity
(``jitter'' ) since $\chi^2_{\rm rv}\approx N_{\rm df}$. The \logg\ value
derived from our MCMC solution is consistent with the \logg\ value from the
analysis of the spectrum, although this is a rather weak constraint because
the \logg\ value from the analysis of the spectrum has a much larger
uncertainty.

\section{Discussion and Conclusions}

 Of the 69 transiting planets currently known with directly measured masses,
only 11 (including WASP-32\,b) have masses greater than 3\,$M_{\rm
Jup}$.\footnote{http://exoplanet.eu} The discovery of WASP-32\,b  will improve
our understanding of how the properties of hot Jupiters vary with the planet's mass.

 \cite{2009Natur.462..189I} claim that, on average, stars with planets
have a lower lithium abundance than normal solar-type stars in the effective
temperature range of 5600\,--\,5900\,K. \cite{2010A&A...512L...5S} claim that
this result was not a consequence of the distribution of age or mass of the
planet host stars. Both these claims have been  disputed by
\cite{2010arXiv1008.0575B} who find that the apparent connection between
lithium abundance and the presence of an exoplanet in the sample of
\citeauthor{2009Natur.462..189I} can be explained by subtle biases in their
sample. \citeauthor{2009Natur.462..189I} did not identify any peculiarities in
the pattern of lithium abundance for planet-host stars with \teff $\ga
5850$\,K. The trend of lithium abundance becomes more complex when these
hotter, more massive stars are considered. The relation between mass and
\teff\ also depends on the age and metalicity of the star so the trend is only
seen clearly if plotted as a function of mass, not \teff.

\begin{figure} 
\plotone{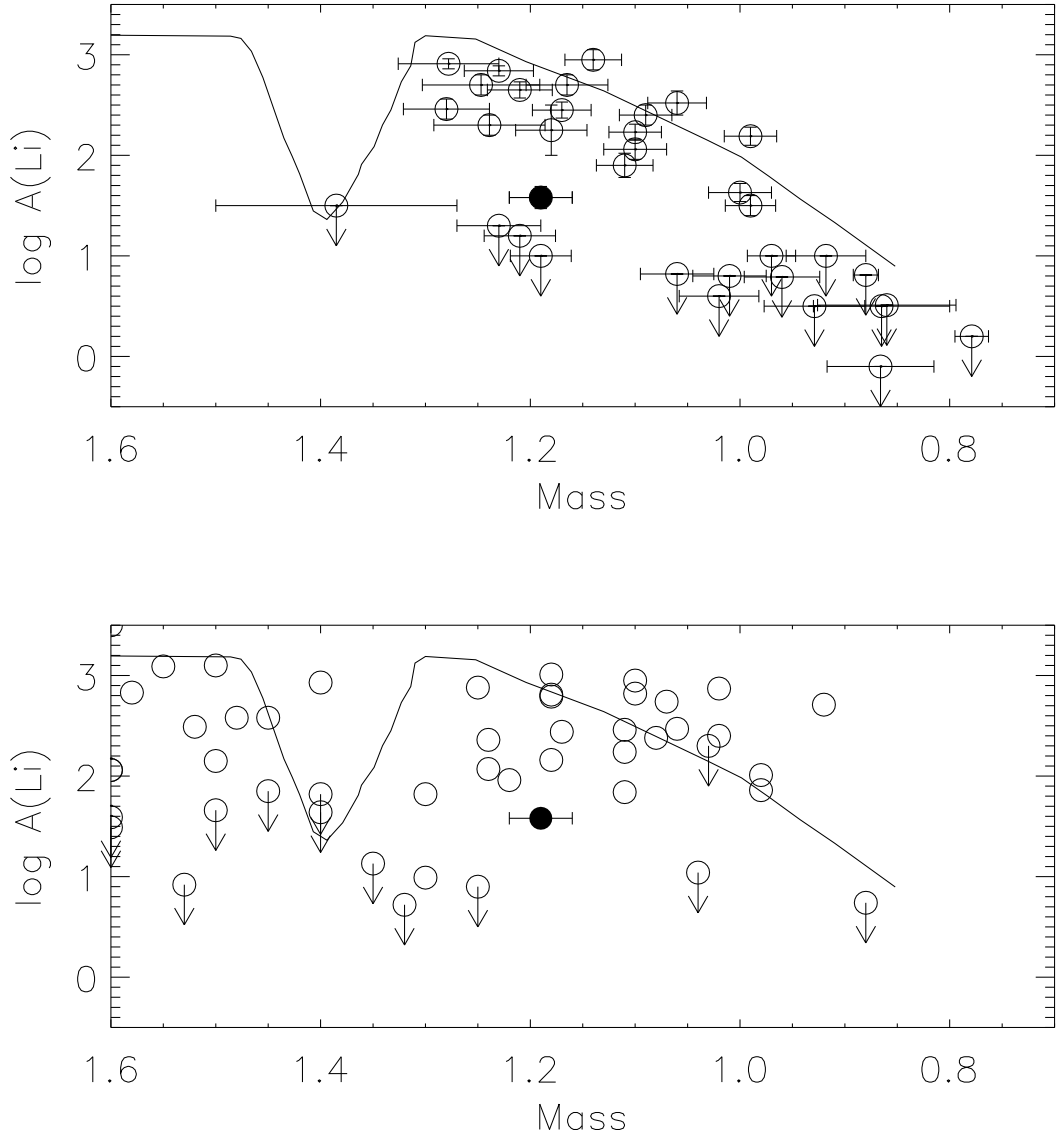} 
\caption{Upper panel:
Lithium abundance as a function of stellar mass for transiting hot Jupiter
planet host stars. Upper limits are indicated with downward pointing arrows.
WASP-32 is plotted with a filled symbol. The solid line is the mean relation
for the Hyades stars from \cite{2004MNRAS.349..757L}. Lower panel: Lithium
abundance for thin-disk stars with [Fe/H]~$>-0.2$ from
\citeauthor{2004MNRAS.349..757L}. \label{MassNLi} } 
\end{figure}

 For field stars with accurate parallaxes, \teff, [Fe/H] and \logg\
measurements, the mass of the star can be estimated by comparison with stellar
models. \cite{2004MNRAS.349..757L} estimated the mass of 451 F-G stars
using this method and compared their surface lithium abundance to similar
stars in various open clusters. Their trend of lithium abundance versus mass
for Hyades stars is shown in  Fig.~\ref{MassNLi} together with their results
for F-G stars with [Fe/H]~$>-0.2$. The location of the prominent dip in the
lithium abundance near masses of 1.4$M_{\sun}$ moves to lower masses for stars
with lower metalicities. The upper limit $\teff\approx 5850\,K$ used by
\cite{2009Natur.462..189I} corresponds to a mass of approximately
1.1$M_{\sun}$. 

 Also shown in Fig.~\ref{MassNLi} are the measured lithium abundances for the
transiting hot Jupiter systems listed in Table~\ref{LiTable}. Stellar masses
for WASP planets in Table~\ref{LiTable} have been re-calculated using the
method of \cite{2010A&A...516A..33E} using the data specified in the reference
provided, unless otherwise noted.  The metalicities of these stars are similar
to the field stars shown in this figure. The tendency suggested by
\cite{2009Natur.462..189I} for planet host stars with masses $\la 1.1M_{\sun}$
to be lithium poor is also seen to be present for transiting hot Jupiter
planets. However, as \cite{2010arXiv1008.0575B} have shown, this tendency may
be a consequence of the  age and metalicity distribution of the samples and be
unconnected to the presence or absence of a planetary companion.

 For more massive hot Jupiter systems there are a few hot Jupiter systems that
are strongly lithium depleted, but the majority of these more massive host
stars have similar lithium abundances to field F-G stars. The position of
WASP-32 in this diagram shows that there is a continuous range of lithium
depletion for planet-host stars with masses of about 1.2$M_{\sun}$. There is
no obvious correlation between the properties of the stars or their planets
and the degree of lithium depletion.  A more detailed analysis of this issue
would benefit from a homogeneous set of age and metalicity estimates for the
host stars of hot Jupiter planets.

\acknowledgements
WASP-South is hosted by the South African Astronomical Observatory and we are
grateful for their ongoing support and assistance. Funding for WASP comes from
consortium universities and from the UK’s Science and Technology Facilities
Council. M. Gillon acknowledges support from the
Belgian Science Policy Office in the form of a Return Grant

\bibliographystyle{pasp}
\bibliography{wasp}

\begin{table} 
\caption{Radial velocity measurements of the star WASP-32. The standard error
of the bisector span measurements, BS, is $2\sigma_{\rm RV}$.} 
\label{rv-data} 
\begin{tabular}{lrrr} 
\hline 
\noalign{\smallskip}
\multicolumn{1}{l}{BJD} & \multicolumn{1}{l}{RV} & 
\multicolumn{1}{l}{$\sigma$$_{\rm RV}$} & \multicolumn{1}{l}{BS}\\ 
--2\,450\,000 & (km s$^{-1}$) & (km s$^{-1}$) & (km s$^{-1}$)\\ 
\noalign{\smallskip}
\hline
\noalign{\smallskip}
5075.8412 & 17.883 & 0.019 & 0.008\\
5117.6744 & 18.922 & 0.123 & 0.088\\
5125.6262 & 18.674 & 0.018 & 0.041\\
5126.6901 & 18.147 & 0.015 & $-0.027$\\
5127.6192 & 17.980 & 0.014 & 0.009\\
5127.7019 & 18.003 & 0.016 & 0.018\\
5128.6424 & 18.761 & 0.015 & 0.024\\
5128.6669 & 18.739 & 0.018 & 0.090\\
5129.5810 & 17.976 & 0.016 & 0.003\\
5129.6984 & 17.866 & 0.016 & $-0.022$\\
5179.5626 & 18.271 & 0.014 & 0.015\\
5180.5828 & 18.641 & 0.012 & 0.052\\
5181.5665 & 17.805 & 0.013 & 0.060\\
5185.5560 & 18.739 & 0.015 & 0.055\\
5188.5509 & 18.773 & 0.019 & 0.063\\
\noalign{\smallskip}
\hline 
\end{tabular} 
\end{table}

\begin{table}[h]
\caption{Stellar parameters of WASP-32 from Spectroscopic Analysis.}
\begin{tabular}{lr}
\hline
\noalign{\smallskip}
Parameter  & \multicolumn{1}{l}{Value} \\ 
\noalign{\smallskip}
\hline
\noalign{\smallskip}
\teff(K)    & 6100 $\pm$ 100 \\
\logg      & 4.4 $\pm$ 0.2 \\
\mictrb(\kms)   & 1.2 $\pm$ 0.1 \\
\vsini(\kms)     & 4.8 $\pm$ 0.8 \\
{[Fe/H]}   &$-$0.13 $\pm$ 0.10 \\
{[Si/H]}   &$-$0.06 $\pm$ 0.10 \\
{[Ca/H]}   &$-$0.02 $\pm$ 0.12 \\
{[Ti/H]}   &$-$0.07 $\pm$ 0.10 \\
{[Cr/H]}   &$-$0.10 $\pm$ 0.09 \\
{[Ni/H]}   &$-$0.16 $\pm$ 0.08 \\
log A(Li)  &   1.58 $\pm$ 0.11 \\
\noalign{\smallskip}
\hline
\end{tabular}
\label{wasp32-params}
\end{table}

\begin{table*} 
\caption{System parameters for WASP-32. The planet equilibrium temperature,
$T_{\rm P}$, is calculated assuming a value for the Bond albedo A$=0$. {\bf
N.B.} an assumed main-sequence mass-radius relation is imposed as an
additional constraint in this solution so the mass and radius of the star are
not independent parameters -- see \citet{2007MNRAS.380.1230C} for details. } 
\label{sys-params} 
\begin{tabular}{lrl} 
\hline 
\noalign{\smallskip}
Parameter (Unit) & \multicolumn{1}{l}{Value} & Notes\\ 
\hline 
\noalign{\smallskip}
$P$ (d) & 2.718659$\pm 0.000008$& Orbital period \\
$T_{\rm c}$ (HJD) & 2455151.0546$\pm 0.0005$ & Time of mid-transit \\
$T_{\rm 14}$ (d) & 0.101 $\pm 0.002 $&Transit duration \\ 
$T_{\rm 12}$ (d) & 0.0161$\pm 0.0015$&Ingress duration \\
$\Delta F=R_{\rm P}^{2}$/R$_{*}^{2}$ & 0.0124$\pm 0.0004$\\
$b=a \cos(i)/$R$_{*}$ & 0.628$\pm 0.004$ & \\
$i$ ($^\circ$) \medskip & 85.3$\pm 0.5$ & Orbital inclination\\
$K_{\rm 1}$ (m s$^{-1}$) & 483  $\pm 6$ & Semi-amplitude of spectroscopic orbit\\
$\gamma$ (m s$^{-1}$) \medskip & 18281$\pm 1$&Radial velocity of barycentre\\
$\sqrt{e\cos\omega}$ & $-0.12 \pm 0.03 $ \\
$\sqrt{e\sin\omega}$ & $ 0.05  \pm 0.07 $\\
$e$ & 0.018 $\pm 0.0065$&Orbital eccentricity\\
$\omega$ & 160 $\pm 30$&Longitude of periastron\\
$M_{\rm *}$ ($M_{\rm \sun}$) & $1.10 \pm 0.03$&Stellar mass\\
$R_{\rm *}$ ($R_{\rm \sun}$) & $1.11 \pm 0.05$&Stellar radius\\
$\log g_{*}$ (cgs) & 4.39$\pm 0.03$ & Logarithmic stellar surface gravity\\
$\rho_{\rm *}$ ($\rho_{\rm \sun}$) \medskip & $0.80 \pm 0.10$ & Mean stellar
density\\
$M_{\rm P}$ ($M_{\rm Jup}$) & 3.60 $\pm 0.07$&Planetary mass\\
$R_{\rm P}$ ($R_{\rm Jup}$) & 1.18 $\pm 0.07$&Planetary radius\\
$\log g_{\rm P}$ (cgs) & 3.77$\pm 0.04$&Logarithmic planetary surface gravity\\
$\rho_{\rm P}$ ($\rho_{\rm J}$) & 2.2 $\pm 0.4$&Mean planetary density\\
$a$ (AU)  & 0.0394$\pm 0.0003$ & Semi-major axis of the orbit\\
$T_{\rm P}$ (K) & 1560$\pm 50$ & Planetary equilibrium temperature\\
\noalign{\smallskip}
\hline 
\end{tabular} 
\end{table*}

\begin{deluxetable}{lrrrr}  
\tablewidth{0pt}
\tablecaption{Lithium abundances and masses for planet host stars.}
\label{LiTable} 
\tablehead{
\colhead{Star} &
\colhead{Mass (M$_{\sun}$)} &
\colhead{log A(Li)}  &
\colhead{Refs.} 
}
\startdata
HD189733    &0.87 $\pm$ 0.05& $< -0.1$        & 1,2 \\  
HD209458    &1.17 $\pm$ 0.04 & 2.7$\pm$0.1     &1,2 \\  
OGLE-TR-10  &1.24 $\pm$ 0.05  & 2.3$\pm$0.1     & 1,2 \\  
OGLE-TR-56  &1.25 $\pm$ 0.06  & 2.7$\pm$0.1     & 1,2 \\  
OGLE-TR-111 &0.86 $\pm$ 0.07  & $<$ 0.51        & 1,2 \\  
OGLE-TR-113 &0.78 $\pm$ 0.02  & $<$ 0.2         & 3,2 \\  
TrES-1      &0.93 $\pm$ 0.05  & $<$0.5          & 1,2 \\  
WASP-1      &1.28 $\pm$ 0.04  & 2.91$\pm$0.05   & 1,4 \\  
WASP-2      &0.88 $\pm$ 0.01  & $<$ 0.81        & 1,5 \\  
WASP-3      &1.22 $\pm$ 0.09 & 2.25$\pm$0.25    &   6 \\  
WASP-4      &0.96 $\pm$ 0.036 & $<$0.79         &   7 \\  
WASP-5      &1.02 $\pm$ 0.038 & $<$0.6          &   7 \\  
WASP-6      &0.87 $\pm$ 0.07  & $<$0.5          &   8 \\  
WASP-7      &1.19 $\pm$ 0.029 & $<$1.0          &   9 \\  
WASP-8      &0.99 $\pm$ 0.024 & 1.5 $\pm$ 0.1   &  10 \\  
WASP-12     &1.28 $\pm$ 0.041 & 2.46 $\pm$ 0.1  &  11 \\  
WASP-13     &1.10 $\pm$ 0.030 & 2.06 $\pm$ 0.1  &  12 \\  
WASP-14     &1.23 $\pm$ 0.033 & 2.84 $\pm$ 0.05 &  13 \\  
WASP-15     &1.21 $\pm$ 0.034 & $<$1.2          &  14 \\  
WASP-16     &1.01 $\pm$ 0.035 & $<$0.8          &  15 \\  
WASP-17     &1.23 $\pm$ 0.040 & $<$1.3          &  16 \\  
WASP-18     &1.21 $\pm$ 0.031 & 2.65 $\pm$ 0.08 &  17 \\  
WASP-19     &0.97 $\pm$ 0.023 & $<$ 1.0         &  18 \\  
WASP-20     &1.09 $\pm$ 0.025 & 2.40 $\pm$ 0.10 &  19 \\  
WASP-21     &0.99 $\pm$ 0.025 & 2.19 $\pm$ 0.09 &  20 \\  
WASP-22     &1.10 $\pm$ 0.025 & 2.23 $\pm$ 0.08 &  21 \\  
WASP-24     &1.17 $\pm$ 0.028 & 2.45 $\pm$ 0.08 &  22 \\  
WASP-25     &1.00 $\pm$ 0.030 & 1.63 $\pm$ 0.09 &  23 \\  
WASP-26     &1.11 $\pm$ 0.027 & 1.90 $\pm$ 0.12 &  24 \\  
WASP-28     &1.06 $\pm$ 0.028 & 2.52 $\pm$ 0.12 &  25 \\  
WASP-30     &1.14 $\pm$ 0.027 & 2.95 $\pm$ 0.10 &  26 \\  
WASP-32     &1.19 $\pm$ 0.030 & 1.58 $\pm$ 0.11 &   \\  
WASP-34     &1.06 $\pm$ 0.035 & $<$0.82         &  27 \\  
TrES-3      &0.92 $\pm$ 0.04  & $< 1.0$         & 28 \\ 
TrES-4      &1.39 $\pm$ 0.10  & $<1.5$          & 28 \\ 
\enddata
\tablerefs{
1. \cite{2009MNRAS.394..272S};
2. \cite{2006A&A...460..251M}; 
3. \cite{2008ApJ...677.1324T};
4. \cite{2006A&A...450..825S};
5. Smalley (priv. comm.);
6. \cite{2008MNRAS.385.1576P};
7. \cite{2009A&A...496..259G};
8. \cite{2009A&A...501..785G};
9. \cite{2009ApJ...690L..89H};
10. \cite{2010A&A...517L...1Q};
11. Smalley (priv. comm.);
12. \cite{2009A&A...502..391S};
13. \cite{2009MNRAS.392.1532J};
14. \cite{2009AJ....137.4834W};
15. \cite{2009ApJ...703..752L};
16. \cite{2010ApJ...709..159A};
17. \cite{2009Natur.460.1098H};
18. \cite{2010ApJ...708..224H};
19. \cite{Pollacco2010};
20. \cite{2010arXiv1006.2605B};
21. \cite{2010arXiv1004.1514M};
22. \cite{2010ApJ...720..337S};
23. \cite{Enoch2010a};
24. \cite{2010arXiv1004.1542S};
25. \cite{West2010};
26. \cite{Anderson2010a};
27. \cite{Smalley2010b};
28. \cite{2009ApJ...691.1145S}
}
\end{deluxetable}

\end{document}